\documentclass[amsmath, amsonfts,twocolumn]{revtex4}

\def\minus{%
  \setbox0=\hbox{-}%
  \vcenter{%
    \hrule width\wd0 height \the\fontdimen8\textfont3
  }%
}

\usepackage{graphicx}
\usepackage{amssymb}
\usepackage{wasysym}
\usepackage{stmaryrd}
 
\usepackage{lineno,hyperref}
\usepackage{amsmath,bm}

\newcommand{\Le}{L_\textup{\tiny e}}
\newcommand{\sig}{{\bm \sigma}}

\newcommand{\Hf}{H_\textup{\tiny f}}
\newcommand{\df}{d_\textup{\tiny f}}
\newcommand{\bvec}{\left(\begin{array}{c}}
\newcommand{\evec}{\end{array}\right)}
\newcommand{\bmat}{\left(\begin{array}{cc}}

\newcommand{\toutin}{\left\{\begin{array}{l}}
\newcommand{\toutind}{\left\{\begin{array}{ll}}
\newcommand{\toutout}{\end{array}\right.}

\renewcommand{\phi}{\varphi}
\newcommand{\phif}{\varphi_\textup{\tiny f}}

\newcommand{\dr}[2]{\frac{\partial #1}{\partial #2}}
\newcommand{\drd}[2]{\frac{\partial^2 #1}{\partial #2 ^2}}

\newcommand{\al}{\alpha}
\newcommand{\dsp}{\displaystyle}
\def\beq{\begin{equation}}
\def\eeq{\end{equation}}
\renewcommand{\div}{\textup{div}}
\newcommand{\grad}{{\bm \nabla}}

\begin{document}

\title{Conversion of Love waves in a forest of trees}
\author{Agn\`es Maurel}
\address{Institut Langevin, ESPCI, CNRS, 1 rue Jussieu, 75005 Paris, France
}
\author{Jean-Jacques Marigo }
\address{Lab. de M\'ecanique des Solides, Ecole Polytechnique, Route de Saclay, 91120 Palaiseau, France
}
\author{S\'ebastien Guenneau}
\address{Aix-Marseille Universit\'e, CNRS, Centrale Marseille, Institut Fresnel, 13013 Marseille, France
}

\begin{abstract} 
We inspect the propagation of shear polarized surface waves akin to Love waves through a forest of trees of same height atop a guiding layer on a soil substrate. We discover that the foliage of trees { brings a radical change in} the nature of the  dispersion relation of these surface waves, which behave like spoof plasmons in the limit of a vanishing guiding layer, and like Love waves in the limit of trees with a vanishing height. When we consider a forest with trees of increasing or decreasing height, this hybrid "Spoof Love" wave is either reflected backwards or converted into a downward propagating bulk wave. 

  \end{abstract}

\maketitle

Research in engineered metasurfaces, which support a host of electromagnetic surface waves, has greatly benefited from the concept of spoof plasmon polaritons that opened new vistas in the microwave regime \cite{pendry}, and this inspired further studies in plasmonics \cite{lalanne} and flat optics \cite{yu}. A particularly appealing object in this area is the so-called rainbow. In their seminal work \cite{tsakmakidis}, Tsakmakidis and co-workers demonstrated that the optical properties of surface electromagnetic waves can be tailored, by varying the surface nanotopology, via surface dispersion engineering. The resulting graded 
metasurface allows for light localization and segregation of different light colors \cite{jang}, a concept which  found a counterpart for sound \cite{zhu}.

\vspace{.3cm}
 In the context of elasticity, previous studies focused on the case of  polarized surface waves known as Rayleigh waves. Despite the differences between these mechanical waves and the electromagnetic waves, it has been shown that Rayleigh waves propagating over elastic crystals share common features with their electromagnetic counterparts, such as the existence of elastic Bragg bandgaps  \cite{achaoui}. These bandgaps have been exploited to create a shielding effect  for Rayleigh waves propagating at $50$ hertz through a soil structured with an array of boreholes  \cite{brule}.  
 Recently,   elastic resonant metasurfaces with  subwavelength structurations  have been considered and the  concept of  rainbow  for optical waves has been translated to Rayleigh waves with the exciting application to the control of  seismic surface waves \cite{colombi1,colombi2,colombi3,colombi4}. The concept {of seismic rainbow} has been first demonstrated for ultrasounds in experiments at the laboratory scale \cite{colombi4}, and extended up    to the geophysical scale \cite{colombi1}. A forest densely populated with trees represents a naturally occurring geophysical metasurface for Rayleigh waves  and an experiment in an actual forest environment \cite{colombi1} confirmed  filtering properties due to the presence of stop bands around $100$ hertz.
The dispersion relation derived in \cite{colombi3} revealed the existence of  an  effective wave that transitions from Rayleigh wave-like to shear wave-like behaviour.  
 These works opened the door to the development of  seismic metasurfaces with the first realization of the so-called meta-wedge that is capable of mode-converting destructive seismic surface waves into mainly harmless downward propagating bulk shear waves.
 
\begin{figure}[h!]
\centering
\includegraphics[width=1\columnwidth]{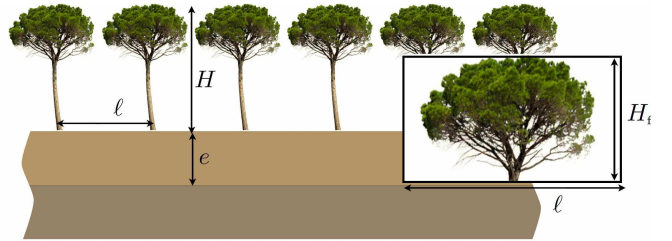}
	\caption{Periodic  array of  trees with spacing $\ell$ and total height $H$; the ground region is surmounted by a guiding layer able to support Love waves. The tree trunks  have a diameter $d$  and  filling fraction $\phi$  and the foliage of height $\Hf$ a surface filling fraction $\phif$. }
	\label{Fig1}
\end{figure}

\vspace{.3cm}
In this Letter, we show that  such seismic metasurfaces can be designed  for  Love waves \cite{love}. Love waves are    shear polarized  surface seismic waves, which  produce a horizontal shaking   particularly deleterious for the foundations of infrastructures. Unlike for Rayleigh waves, Love waves require a guiding layer to propagate at the air-soil surface and we shall see that these surface waves  are particularly sensitive to the shape of structural elements above the soil, in the present case,  a forest of trees with some foliage {(Fig. \ref{Fig1})}.  The effective dispersion relation reflects a cooperation between the guiding layer and the trees,  {resulting in} effective bandgaps for a hybrid wave. 
As a result it is shown that  a forest of trees with varying  height can reflect, localize or convert a Love wave.

\vspace{.3cm}

Considering a coordinate system $(x,y,z)$, the propagation of elastodynamic waves in a heterogeneous isotropic medium is governed by the Navier equation 
\beq
\dsp \rho\drd{\bf u}{t}=\rm{div}\sig,   \quad
\sig={\bf C}:\nabla{\bf u}, 
\eeq
with $t$ the time variable. The body force is assumed to be zero, ${\bf C}=(C_{ijkl}), $ is  the 4-order elasticity tensor, which satisfies Hookes' law $C_{ijkl}=\lambda\delta_{ij}\delta_{kl}+\mu(\delta_{ik}\delta_{jl}+\delta_{il}\delta_{jk})$, $i,j,k,l=1,2,3$, with $\lambda$ and $\mu$ the Lam\'e coefficients, $\rho$ {is} the density and {$({\bf u},\sig)$ the vector displacement field and the stress tensor, respectively}.
 For an elastic medium which is invariant along one direction, say $y$, the Navier equation splits into an in-plane equation on ${\bf u}=(u_x,u_z)$  and  an out-of-plane equation on $u=u_y$. {We focus on the out-of-plane polarization which concerns  
 Love waves; if} one further assumes some time-harmonic dependence $e^{-i\omega t}$, with $\omega$ the angular wave frequency,  this problem takes the simple form
\beq\label{eqshear}
 \rm{div}\sig+\rho\omega^2 u=0, \quad \sig=\mu \grad u. 
\eeq

\begin{figure}[h!]
\centering 
\includegraphics[width=.8\columnwidth]{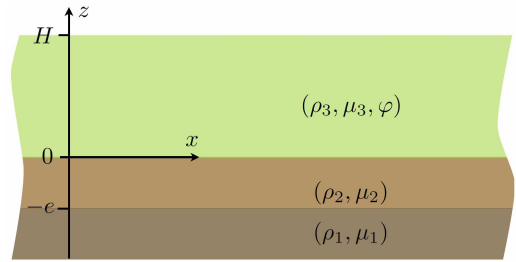}
	\caption{The effective problem, in which the region of the trees $0< z<H$ has been replaced by an equivalent (anisotropic) layer where  the homogenized equation \eqref{eq1} applies.}
	\label{Fig2}
\end{figure}

Let us now dive into our problem. We consider the two-dimensional configuration of a forest of trees  periodically spaced by a distance $\ell$; the ground is composed of a  layer  with a lower velocity than that of the soil substrate (Fig. \ref{Fig1}).  In the absence of trees, this low velocity  layer can support Love waves, which propagate  within the layer and vanish when moving far from it in the substrate.  We shall see that   it is possible to describe how the guiding layer couples to the trees to produce a new type of guided waves, that we term "Spoof Love" waves.  
 Using   homogenization tools,  the region of the trees can be replaced by an equivalent slab filled with a homogeneous anisotropic medium \cite{mercier,marigoSIAP}, {see also the [Supplemental Material]}. In this medium, $(u,\sig)$ satisfy 
\beq
\dsp \sig=\mu_3 \bmat 0 &0 \\[10pt] 0 &\phi \evec\\ \grad u, \quad \div \sig +\rho_3 \omega^2 \phi \, u=0,\label{eq1}
\eeq
with $(\rho_3,\mu_3)$ the mass density and the shear modulus of the wood which composes the trees, and $\phi$ the filling fraction of tree trunks {(denoting $d$ the diameter of the trunk, $\phi=d/\ell$)}.  It is worth noting that \eqref{eq1} tells us {that} the propagation is allowed along the trees only, with the  wavenumber $\omega/c_3$, $c_3=\sqrt{\mu_3/\rho_3}$,  as in a single tree (since we have $\partial_{zz}{u}+(\omega^2/c_3^2)\, u=0$).   
In the complete homogenized problem shown in Fig. \ref{Fig2}, boundary conditions have to be applied at the interfaces of the effective slab at $z=0$ and $z=H$. 
At the interface $z=0$ with the guiding layer, the boundary conditions are the   usual continuity relations of $u$ and {$\sigma_z$}.  At  the interface 
$z=H$ with the air above the trees,  a non intuitive effective condition applies of the form
\beq\label{bc}
\sigma_z= -\Le \dr{\sigma_z}{z},\quad \Le=\Hf \left(\frac{\phif}{\phi}-1\right),\quad \textup{at}\; z=H,
\eeq  
with $\Hf$ the {foliage} height and $\phif$ the {foliage} filling fraction{, both of} which characterize the foliage (inset in Fig. \ref{Fig1}). Finally, the propagation is described by \eqref{eqshear} in the guiding layer, for $0>z>-e$, with the material parameters $(\rho_2,\mu_2)$ and in the substrate, for $z<-e$, with the material parameters $(\rho_1,\mu_1)$.

We start our analysis in the absence of foliage, whence \eqref{bc} reduces to ${\sigma_z}_{|z=H}=0$. 
Looking for a guided wave solution, {which is the solution of the problem} in the absence of source, we can derive the dispersion relation. Specifically, we are looking for solutions of the form
\beq\label{sol}
u(x,z)=\toutind
e^{\al_1(z+e)}e^{i\beta x}, & z< -e,\\
\dsp \left[A\cos k_2z+B \sin k_2z \right]e^{i\beta x}, & -e<z<0,\\
\dsp C\,\cos k_3(z-H)\;e^{i\beta x}, & 0<z<H.
\toutout
\eeq
In \eqref{sol}, we have accounted for the   condition  ${\sigma_z}_{|z=H}=0$, and we have defined the vertical wavenumbers
\beq\label{al}
\al_1=\sqrt{\beta^2-\frac{\omega^2}{c_1^2}}, \quad k_2=\sqrt{\frac{\omega^2}{c_2^2}-\beta^2},\quad  k_3=\frac{\omega}{c_3}.
\eeq
We are looking for $\al_1$ real positive and  $\beta$ real, which correspond to a guided wave in the ground, but $k_2$ can be {\em a priori} real or imaginary; if $k_2$ is imaginary, the wave is evanescent in the guiding layer as it is in the ground.
Applying the continuity of $u$ and $\sigma_z$ at the interfaces at  $z=-e$ and 0,
\eqref{sol}   leaves us with 4 equations on $(A,B,C,\beta)$ for each frequency $\omega$, from which the dispersion relation $\beta(\omega)$ can be {inferred}. It reads as 
\beq\label{disp1}
1-\frac{\mu_2 k_2}{\mu_1\al_1}\tan k_2e-\frac{\mu_3    k_3}{\mu_2k_2}\phi\tan k_3H\left(\tan k_2e+\frac{\mu_2 k_2}{\mu_1\al_1}\right)=0.
\eeq
The above dispersion relation describes {a} guided wave supported by the guiding layer coupled to the trees. Obviously for $H=0$, we recover the dispersion relation of Love waves
$\alpha_1=(\mu_2/\mu_1) k_2 \tan k_2e$ \cite{love}. An other interesting limit is  $e=0$ where we see that   the trees alone are able to support guided waves, with $\alpha_1=(\mu_3 /\mu_1) \phi k_3 \tan k_3H$; in the case $\mu_3 =\mu_1$, we recover the dispersion relation of the so-called spoof plasmon corresponding in acoustics and electromagnetism to 
{guided} waves propagating over a rough rigid surface \cite{pendry}. 
To account  for the foliage at the tree top, it is sufficient to modify in \eqref{sol} the form of the solution for $0<z<H$, specifically  
\beq\label{sol2}
u(x,z)=C \left[\cos k_3(z-H)+k_3\Le\sin k_3(z-H)\right]e^{i\beta x},
\eeq
which satisfies \eqref{bc}, and  eventually the complete dispersion relation is obtained in the form
\beq\label{disp2}
F_1\left[1-\frac{\mu_2 k_2}{\mu_1\al_1}\tan k_2e\right]-F_2\;\frac{\mu_3    k_3}{\mu_2k_2}\phi \left(\tan k_2e+\frac{\mu_2 k_2}{\mu_1\al_1}\right)=0,
\eeq
with 
\beq\toutin
\dsp F_1=1-k_3\Le\tan k_3H, 
\\ \dsp F_2=\tan k_3H+k_3\Le.
\toutout\eeq
Obviously, for $\Le=0$, \eqref{disp2} simplifies in \eqref{disp1}.
 
 \vspace{.3cm}
 
 From now on, we use the following material parameters:  $\rho_1=1300$ kg.m$^{-3}$, $c_1=495$ m.s$^{-1}$ for the ground,  $\rho_2=2600$ kg.m$^{-3}$, $c_2=350$ m$s^{-1}$ for the guiding layer, and $\rho_3=450$ kg.m$^{-3}$, $c_3=1200$ m.s$^{-1}$ for the wood. The dimensions are $e=2$ m, $H=10$ m, $d=0.3$ m, $\ell=2$ m ($\phi=0.15$); when the foliage is considered, we use   $\df=1.5$ m ($\phif=0.75$) and $\Hf=1$ m, thus the height of trunk is 9 m in this case (note we assume same elastic parameters for trunk and foliage, but this assumption can be lifted [Supplemental Material]). 
 \begin{figure}[h!]
\centering
\includegraphics[width=1\columnwidth]{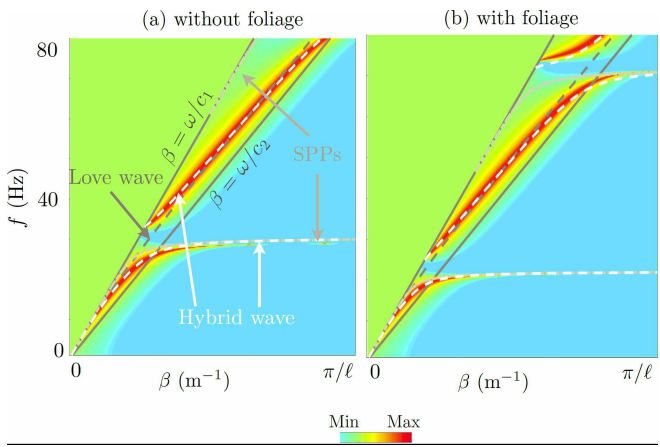}
	\caption{Dispersion relation of the hybrid "spoof Love" wave for $H$= 10 m, calculated numerically by means of the divergence of the reflection coefficient $|R|$ (in log, color scale)  and the dispersion relation \eqref{disp1}, \eqref{disp2} (dashed white lines). The hybrid wave results from a cooperation between the guiding layer supporting Love wave (dashed dark grey lines) and the trees supporting spoof plasmons (SPPs, light grey lines); the light lines $\beta=\omega/c_1, \omega/c_2$ are  reported in plain grey lines.}
	\label{Fig3}
\end{figure} 
We begin our physical discussion of Spoof Love waves with the inspection of their dispersion relation  in the actual problem by computing numerically the reflection coefficient $R$ for an incident evanescent wave (using a multimodal method based on eigenfunction expansions \cite{maurel}).
 Specifically   we consider a solution for $z<-e$ of the form 
$u(x,z)=e^{i\beta x}(e^{-\al_1z}+Re^{\al_1(z+e)})$,
with  $\al_1$ in \eqref{al}. Above the light line $\beta<\omega/c_1$,  $\al_1$ is purely real whence $|R|=1$; but below the light line   $\al_1$ is purely imaginary and $|R|$ is unbounded; when   $|R|=\infty$ we recover  a guided wave as in \eqref{sol}. Results on the reflection coefficient $R$  are reported in Fig. \ref{Fig3}. For comparison, we report the dispersion relations of the Love waves in the layer on its own and those of the spoof plasmons  in the trees on their own. The actual dispersion relations show that the guiding layer couples to the trees, resulting in a hybrid guided wave, which appears to be accurately described by our model \eqref{disp1} and \eqref{disp2}. Roughly speaking, the  Spoof Love wave remains close to the Love wave except in the vicinities of the cut-off frequencies of the spoof plasmons  corresponding to the resonances of a single tree $f_\textup{\tiny c}=(2n+1)c_3/(4H)$, $n$ integer (the  first cut off frequency  around  $30$ Hz  is {clearly seen}). There, the trees dominate and the wave becomes evanescent not only in the soil substrate but also in the guiding layer, with $\beta> \omega/c_2$. 
  When the foliage is accounted for,  the cut-off frequencies $f_\textup{\tiny c}$ are significantly decreased, a fact that is already true for the spoof plasmons in the absence of guiding layer ($e=0$ in \eqref{disp2}).  It is worth noting that such a  decrease in the resonance frequency  is not attributable only to a larger  inertia of the tree; for instance,  increasing $\phi$ does not affect the asymptotes at   $f_\textup{\tiny c}$, which remains dictated by $H$ only.

\begin{figure}[h!]
\centering
\includegraphics[width=1\columnwidth]{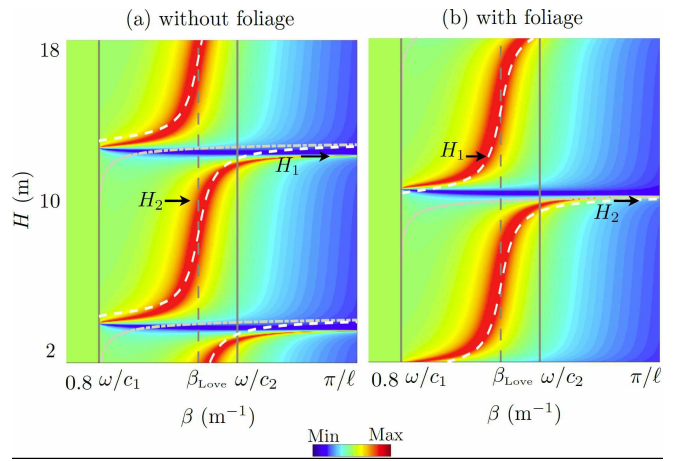}
	\caption{ Dispersion relation of the spoof Love waves in the plane ($\beta,H)$   at $f=70$ Hz;  (a) without  and (b) with foliage  (same representation as in Figs. \ref{Fig3}). $\omega/c_1=0.89$ m$^{-1}$, $\beta_\textup{\tiny Love}=1.15$ m$^{-1}$, $\omega/c_2=1.25$ m$^{-1}$, $\pi/\ell$=1.57 m$^{-1}$. 
	}
	\label{Fig4}  
\end{figure}

\begin{figure}[h!]
\centering
\includegraphics[width=.8\columnwidth]{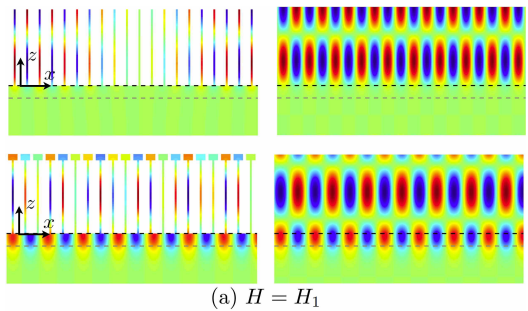}
	\caption{Left panels: displacement fields of the Spoof Love waves without and with foliage  at $f=70$ Hz and (a) $H=H_1$=12.2 m, (b) $H=H_2$= 10.2 m (see black arrows in Fig. \ref{Fig4}). The right panels show the corresponding displacement fields of the guided waves in the effective problem,  \eqref{sol}, \eqref{disp1}  and \eqref{sol2}, \eqref{disp2}. }
	\label{Fig5}
\end{figure}

 We now turn to the influence of the tree height. 
The Figs. \ref{Fig4} show the dispersion relations for $H\in (2;18)$ m  at the frequency $70$ Hz.   The main features observed in Figs. \ref{Fig3} are recovered. 
At this frequency,  resonances occur for trees of heights  $H\simeq 4$ m and  $13$ m, resulting in two bandgaps for  $H\in (3.5,4)$ m and   $H\in(12.3,13)$ m  respectively. With the foliage, the first gap  is shifted  to  $H\in(10,10.5)$ m (and the second to $H\in (1.5,1.9)$ m).  
We also reported in  Fig. \ref{Fig5}(a) the  displacement fields   of the spoof Love waves at the upper limit of the  bandgaps ($\beta=1.5$ m$^{-1}$, close to $\pi/\ell$)  for a forest of trees without foliage and $H_1=12.2$ m (see black arrow $H=H_1$ in Fig. \ref{Fig4}(a)). When the trees  support a foliage, the guided wave changes significantly its shape, being now close to a classical Love wave with $\beta\simeq 1$ m$^{-1}$. Reversely, a forest of trees of height  $H=H_2=10.2$ m with foliage supports a highly confined guided wave with $\beta=1.5$ m$^{-1}$  while suppressing the foliage produces a drastic change in the characteristics of the guided waves (black arrows $H=H_2$ in Figs. \ref{Fig4}). 
In the 4 considered cases, we report for comparison the fields given by \eqref{sol} along with \eqref{disp1}  (and \eqref{sol2} with \eqref{disp2}), which confirms the capacity of the model to predict the main features of the Spoof Love wave.

 \vspace{.3cm}

Beyond the important effect of the foliage, another interesting feature is visible, which may impact significantly the propagation of guided waves through a forest of trees with increasing or decreasing heights. Let us assume that a local analysis can be done, which means that the wavenumber $\beta(H)$ in a forest of trees with varying heights can be estimated from our present analysis, and consider the cut-off frequency for $H=4$ m in Fig. \ref{Fig4}(a). When the wave propagates from shorter $H<$ 4 m towards higher  trees, the local wavenumber $\beta(H)$ increases along the branch $\beta(H<$ 4 m). Eventually, it reaches the value $\pi/\ell$ where the confinement in the trees is maximum (the wave is already evanescent in the guiding layer). On the contrary, if the wave propagates from higher to shorter height trees, $\beta(H)$ is  decreasing along  the branch $\beta(4<H<13$ m). In this case,  it reaches the wavenumber $\beta_1=\omega/c_1$ of the shear wave in the bulk  where the confinement vanishes. Owing to this analysis, it seems not too hazardous to state that a wave propagating along trees with decreasing height  becomes more and more adapted to be converted into a shear wave in the bulk.
{Conversely, the wave propagating along trees with increasing heights is more and more confined being eventually supported by the trees only. When reaching the tree realizing the resonance, it is not adapted to be converted in a bulk wave, nor adapted to pass through the cut off frequency; as an alternative, it is simply reflected backwards.  This strongly non symmetric behavior of the spoof Love wave, also observed for Rayleigh waves with the metawedge \cite{colombi2}, is illustrated in Fig. \ref{Fig6}. We considered a point source outside a forest of 29 trees with heights varying from 16 to 2 m (with a decrement of 0.5 m) which generates a Love wave in the guiding layer. 
 \begin{figure}[h!]
\centering
\includegraphics[width=.8\columnwidth]{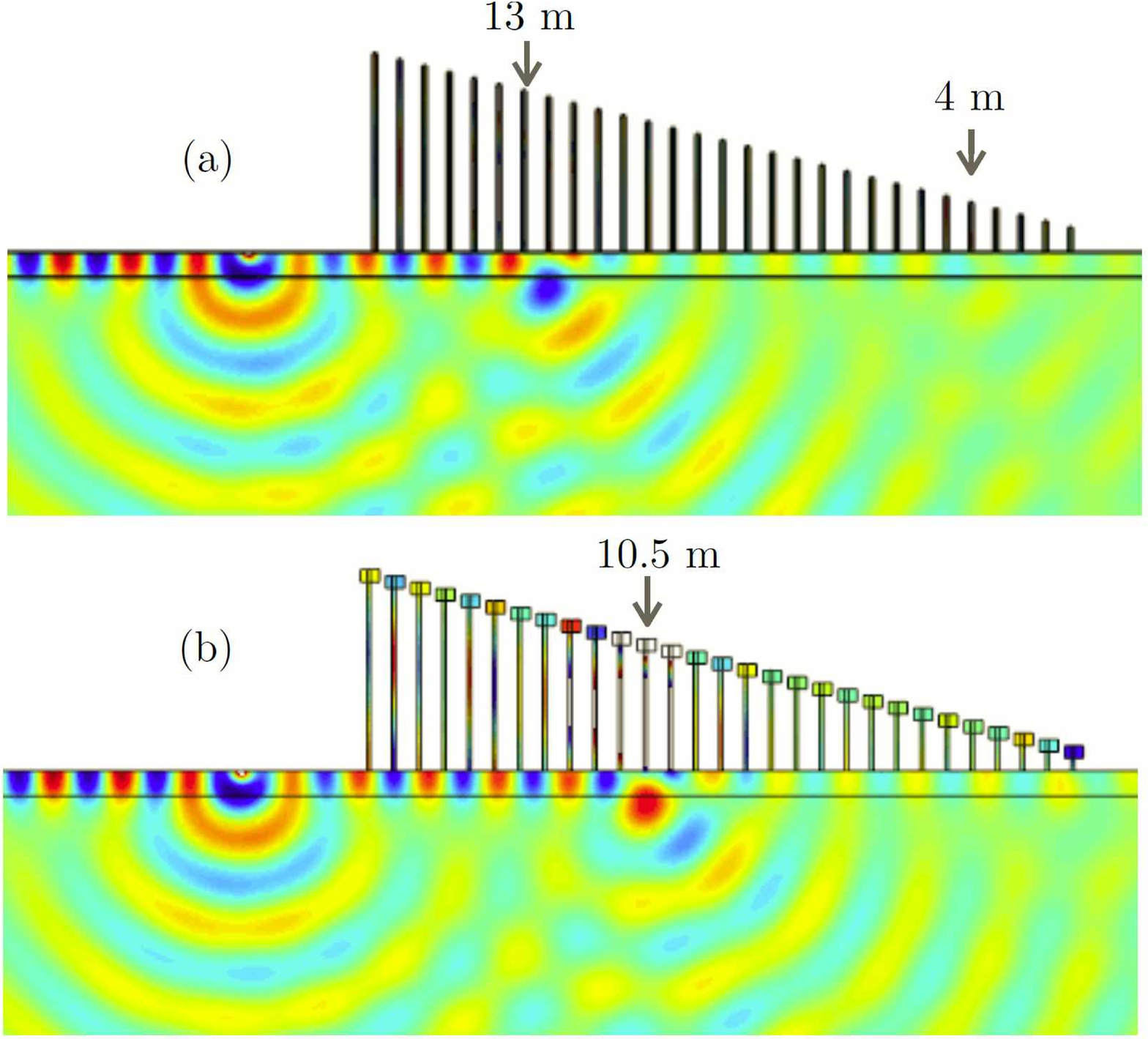}
\includegraphics[width=.8\columnwidth]{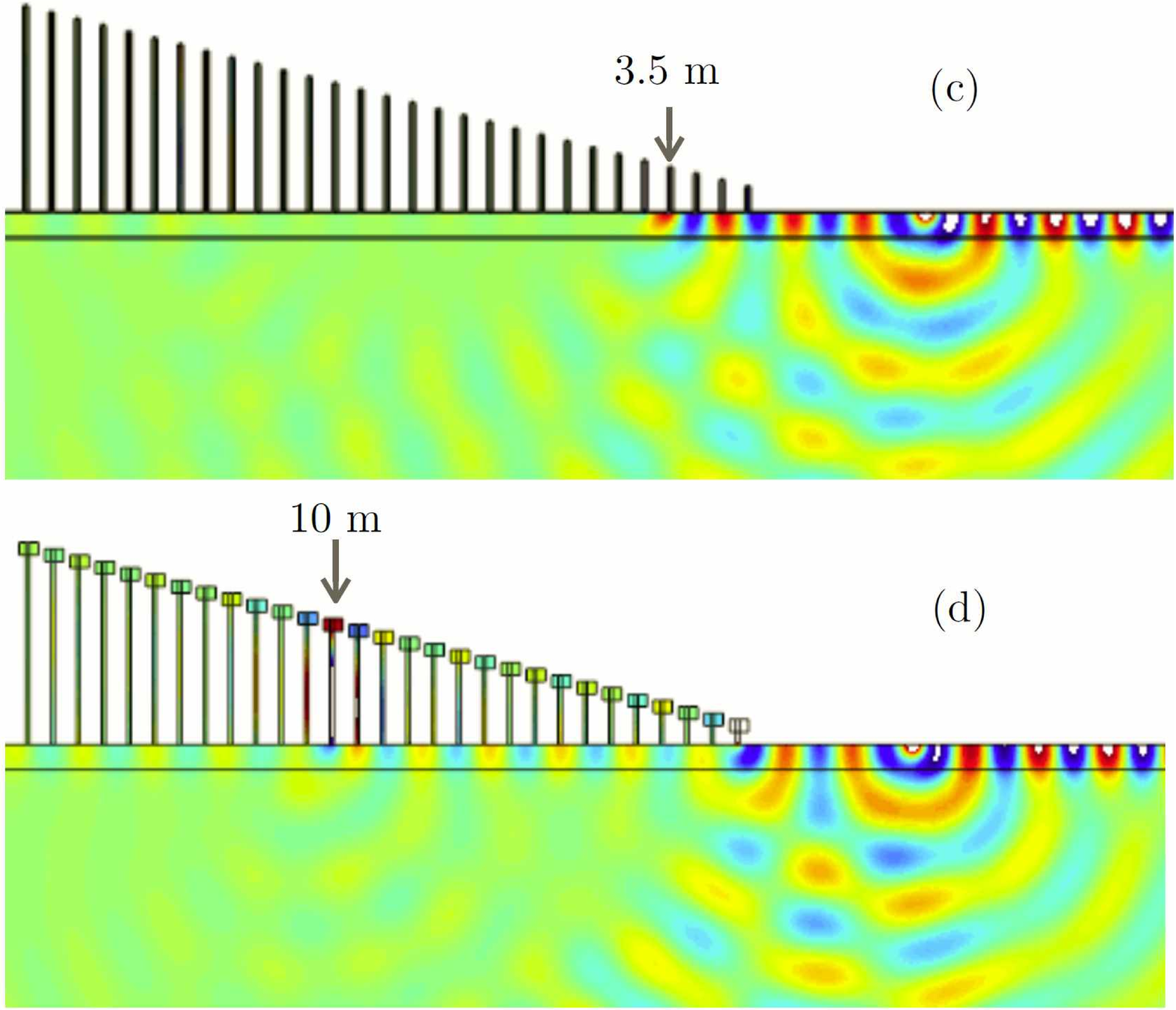}
	\caption{Interaction of a Love wave with a forest consisting of $29$ trees with the same diameter, foliage and spacing as in Fig. \ref{Fig5}, with heights varying from $16$ m to $2$ m.; Linear color range is in arbitrary units (white color is out of scale). }
	\label{Fig6}
\end{figure}
	When the Love wave is incident from the taller edge of the forest (a,b), it is converted into a Spoof Love wave that propagates within the forest  until it reaches a resonant tree; there,  it is converted into a downward propagating shear-polarized bulk wave. These "turning points"    coincide reasonably well with  the upper limits of the bandgaps reported in Figs. \ref{Fig4}, at 13 and 4 m for trees without foliage and at 10.5 m for trees with foliage.  In contrast,  when the Love wave is incident from the shorter edge of the forest, it is converted into a spoof Love wave which eventually is trapped for some tree height, resulting in a significant  back-reflection (we reported the heights corresponding to the upper limits of the bandgaps in Figs. \ref{Fig4}).
The "aspiration" effect of the wave by the trees near the turning points is visible; it is  consistent with our findings that the spoof Love wave becomes evanescent in the guiding layer near the upper limit of the bandgap ($\beta>\omega/c_2$).

In this Letter, we have reported  direct numerical observations of   shear surface waves propagating in a forest of trees atop a guiding layer. Analytical dispersion relation shows that this wave shares common features with both Love waves and spoof plasmons, whose dispersion relations are recovered in two limit cases (trees of vanishing height, and guiding layer of vanishing thickness, respectively), hence the nickname 'spoof Love wave'. 
The dispersion relation reveals bandgaps dictated by the local resonances of a single tree, with  strongly asymmetric characteristics of the wave at the band edges. This finding shows  that spoof  Love waves propagate  following  the same scenario than the one  analyzed for Rayleigh waves. When  propagating in a forest with increasing height, they are easily converted into a downward bulk wave, while when propagating in a forest with decreasing tree height, they are  eventually  aspired within the trees resulting in a strong backscattering due to a rainbow effect. 
Finally, we have shown   that the presence of foliage significantly affects the local resonances of the trees, hence the whole dispersion relation, a phenomenon which is neatly captured by our model.
Interesting extensions include the study of 
other boundary layer effects due to  heterogeneities as the tree roots or the presence of rocks in the guiding layer and the  three dimensional analysis.

It is worth noting that  our analysis applies mutatis mutandis to Love waves interacting with taller resonators resulting in  lower frequency bandgaps. This opens potential applications  in civil engineering where wind farms  (with wind turbine of 50-100 m height) could be used for low frequency   filtering (1 to 10 Hz), or in the study of  site-city interaction involving tall buildings in sedimentary basins.


\begin{thebibliography}{1}

\bibitem{pendry} 
Pendry JB, Martin-Moreno L, Garcia-Vidal F 2004, Mimicking surface plasmons with structured surfaces. Science 305, 847-848

\bibitem{lalanne} 
Lalanne P, Hugonin JP 2006, Interaction between optical nano-objects at metallo-dielectric interfaces, Nature Physics 2 (8), 551

\bibitem{yu}
Yu N, Capasso F 2014, Flat optics with designer metasurfaces, Nature Materials 13, 139-149

\bibitem{tsakmakidis}
Tsakmakidis KL,  Boardman AD, Hess O 2007 Trapped  rainbow  storage  of  light  in  metamaterials, Nature 450, 397-401

\bibitem{jang}
Jang M, Atwater H 2011 Plasmonic  rainbow  trapping  structures  for  light  localization  and  spectrum  splitting, Phys. Rev. Lett. 107, 207401

\bibitem{zhu}
Zhu J, Chen Y, Zhu X, Garcia-Vidal FJ, Yin X, Zhang X, Zhang X 2013, Acoustic rainbow trapping, Sci Rep 3, 1728



\bibitem{achaoui}
Achaoui Y, Khelif A, Benchabane S, Robert L, Laude V 2011 Experimental observation of locally-resonant and bragg band gaps for surface guided waves in a phononic crystal of pillars. Phys. Rev. B 83, 10401

\bibitem{brule}
Brule S, Javelaud EH, Enoch S, Guenneau S 2014 Experiments on seismic metamaterials:  Molding surface waves, Phys. Rev. Lett. 112, 133901

\bibitem{colombi4}
Colombi A, Ageeva V, Smith JR, Clare A, Patel R, Clark M, Colquitt DJ, Roux P, Guenneau SRL, Craster RV. 2017 Enhanced sensing and conversion of ultrasonic Rayleigh waves by elastic metasurfaces. Sci Rep 7, 6750

%
\bibitem{colombi2}
Colombi A, Colquitt, DJ, Roux P, Guenneau S, Craster RV. 2016 A seismic metamaterial: The resonant metawedge. Sci Rep 6, 27717.


\bibitem{colombi1}
Colombi A, Roux P, Guenneau S, Gueguen P, Craster RV 2016, Forests as a natural seismic metamaterial: Rayleigh wave bandgaps induced by local resonances, Sci Rep 6, 19238


\bibitem{colombi3}
Colquitt DJ, Colombi A, Craster RV, Roux P, Guenneau S 2017 Seismic metasurfaces: Sub-wavelength resonators and Rayleigh wave interaction. J Mech Phys Solids 99, 379-393. 


\bibitem{love} 
Love AEH 1911, Some problems of geodynamics (Cambridge University Press)


\bibitem{mercier}
Mercier JF, Cordero ML, F\'elix S, Ourir A,  Maurel A 2015, Classical homogenization to analyse the dispersion relations of spoof plasmons with geometrical and compositional effects. Proc. R. Soc. A  471, 2182,  20150472.


\bibitem{marigoSIAP} 
Marigo JJ,  Maurel A 2017, Second order homogenization of subwavelength stratified media including finite size effect. SIAM J. Appl. Math. 77(2), 721-743.

\bibitem{maurel}
Maurel A, Mercier JF,  F\'elix S 2014, Wave propagation through penetrable scatterers in a waveguide and through a penetrable grating.  J.  Acoust. Soc. Am. 135(1), 165-174.

%
%
%
%


\end{thebibliography}
\end{document}